\documentclass[paper]{ieice}
\usepackage[dvips]{graphicx}

\def\argmin{\mathop{\mbox{argmin}}}

\field{A}
\vol{87}
\no{10}
\SpecialSection{Information Theory and Its Applications}
\title{Two Methods for Decreasing the Computational Complexity of the MIMO ML Decoder}
\authorlist{% fill arguments of \authorentry, otherwise error will be caused. 
\authorentry{Takayuki FUKATANI}{n}{ue}
\authorentry[ryutaroh@it.ss.titech.ac.jp]{Ryutaroh MATSUMOTO}{m}{ue}
\authorentry{Tomohiko UYEMATSU}{m}{ue}
}
\affiliate[ue]{The author are with the Department of Communications and
Integrated Systems, Tokyo Institute of Technology, Tokyo 152-8552 Japan.}
\received{2004}{1}{16}
\revised{2004}{4}{14}
\finalreceived{2004}{6}{7}

\begin{document}
\maketitle
\begin{summary}
We propose use of QR factorization with sort and Dijkstra's algorithm for decreasing the computational complexity of
the sphere decoder that is used for ML detection of signals on the multi-antenna
fading channel. QR factorization with sort  decreases the complexity of 
searching part of the decoder with small increase in the complexity
required for preprocessing part of the decoder.
Dijkstra's algorithm  decreases the complexity of
searching part of the  decoder with increase in the storage complexity.
The computer simulation demonstrates that the complexity of the
decoder is reduced by the proposed methods significantly.
\end{summary}

\section{Introduction}

In the multi-antenna mobile communication, it is well-known that use of multiple transmit
and receive antennas linearly increases the channel capacity of a
frequency nonselective fading channel 
with the channel state
information (CSI) known at the receiver \cite{Telatar,Foschini}. In the
case of the uncoded multi-antenna systems, the computational complexity
of the naive maximum-likelihood (ML) decoding algorithm grows exponentially with the number
of transmit antennas, so we need an efficient algorithm to implement ML
decoding. On the multi-antenna fading channel, if
the receiver has CSI, the receiver can compute the set of ideal received
signal points considering only influence of the fading
and disregarding influence of the additive noise. So when the noise at each
receive antenna is the additive white Gaussian, to implement ML
decoding we search for the ideal received signal point closest to
the actual received signal point. By regarding the ideal received points
as lattice points, the ML decoding problem is reduced to
the classical closest lattice point search problem. Fincke and Pohst proposed
an efficient algorithm for that problem \cite{Pohst}, and recently it was applied to the decoding problem and  called
sphere decoder (SD) \cite{Universal}.  

{
SD can be divided into the two parts. The first part computes QR factorization (or Cholesky factorization) of the fading matrix. The second part computes the ML estimate of transmitted signal from the received signal and QR factorization. We call the first part preprocessing part and the second part searching part. 
}
In this paper we propose QR factorization with sort and use of
Dijkstra's algorithm for decreasing the computational complexity of SD.  QR factorization with sort gives an efficient order of decisions on signal components.
It reduces the complexity of {searching part} with increase in the complexity of preprocessing part.
Dijkstra's algorithm is an efficient algorithm used to solve the  shortest path problem in the
graph. We apply this algorithm to searching part.
It reduces the complexity of {searching part} with increase in the storage complexity. 

{
The QR factorization with sort modifies only preprocessing part and use of Dijkstra's algorithm modifies only searching part. Thus these improvements are independent and can be used together or alone.
}

{
This paper is organized as follows: Section 2 introduces the channel
model of  the multi-antenna fading channel and shows how the
original SD works. Section 3 introduces QR factorization with sort and Section 4
proposes application of Dijkstra's algorithm to SD. Section 5 shows the comparison between
the complexity of the original SD and that of SD using the proposed methods by the
computer simulations. These simulations show that the proposed methods
decrease the complexity of a decoder significantly.
}
%%%%%%%%%%%%%%%%%%%%%%%%%%%%%%%%%5
\section{Original sphere decoder}
\subsection{Channel model}
Suppose that we have the uncoded system with $t$
transmit antennas and $r$ receive antennas, that the noise at each
receive antenna is the additive white Gaussian, and that the receiver
has CSI. At the transmitter, information sources are
demultiplexed into $t$ substreams, and transmitted by transmit
antennas. {Let $\mathbf{a}$ be a $(t \times 1)$ vector consisting of complex
envelopes of transmitted signals with the signal constellation
$\mathbf{S}$, $M$ the $(r \times t)$ fading matrix} whose
($k$,$j$) entry is a complex fading coefficient between $j$-th
transmit antenna and $k$-th receive antenna, {\boldmath{$\nu$}} 
a $(r \times 1)$ complex vector whose component is noise at each
receive  antenna, and $\mathbf{x}$ 
a $(r \times 1)$ complex vector whose component is  the received signal component at each
receive antenna. The model of this channel is written as
\begin{equation}
\mathbf{x}=M\mathbf{a}+\mbox{\boldmath{$\nu$}} \qquad (\mathbf{a}\in
\mathbf{S}^t)\label{multi}.
\end{equation}
On the channel described by Eq.\ (\ref{multi}), when the components of
{\boldmath{$\nu$}} are independent complex Gaussian random variables, the ML decoding
problem can
be reduced to the closest lattice point search problem
for the set of lattice points $ \{M\mathbf{a} \mid \mathbf{a}\in
\mathbf{S}^t \}$ and a received signal point
$\mathbf{x}$. {See \cite{Universal,radius} for details.}

We also remark that when we use linear space-time coding, the ML decoding problem is
reduced to the closest lattice point search problem by
describing the channel as  Eq.\ (\ref{multi}) \cite{MIMO}.

%%%%%%%%%%%%%%%%%%%%%%%%%%%%%%%%%%%%%%%%%%%%
\subsection{Algorithm}
In this section, we show how the original SD works when $t\leq r$. Fincke and Pohst's original method treats real numbers, and we can
treat complex numbers in the almost same way \cite{radius}. 

To implement ML decoding on the channel described by Eq.\ (\ref{multi}), we
must compute the ML transmit signal
$\mathbf{\hat{a}}=(\hat{a}_1,\cdots ,\hat{a}_t)^T$ equal to
\begin{equation}
\argmin_{\mathbf{\hat{a}} \in \mathbf{S}^t}\| \mathbf{x} - M \mathbf{\hat{a}} \|. \label{min}
\end{equation}
To compute Eq.\ (\ref{min}), SD considers a
sphere with center at the received signal in the complex Euclidean space. If there are lattice points in the sphere, the closest point is in the sphere. SD takes a suitable value as the radius of sphere, and searches for 
lattice points in the sphere.   

First we compute QR factorization of $M$  and obtain an
upper triangular matrix $R$ and a unitary matrix $Q$ with $M=QR$. Since $Q$ is a unitary
matrix,   
\begin{equation}
\| \mathbf{x} - M \mathbf{\hat{a}} \| ^2 = \| Q^* \mathbf{x} - Q^* M \mathbf{\hat{a}} \| ^2 = \|Q^* \mathbf{x} - R\mathbf{\hat{a}} \| ^2 .
\end{equation}
Let $\mbox{\boldmath $\rho$}= Q^* \mathbf{x}=(\rho_1,\cdots, \rho_r)^T$, $C$ the square
of suitable radius {and $r_{ij}$ the $(i,j)$ element of $R$}. The lattice points $M \mathbf{\hat{a}}$ that satisfy 
\begin{eqnarray}
\| \mathbf{x} - M \mathbf{\hat{a}} \| ^2&=&\|R\mathbf{\hat{a}} -\mbox{\boldmath $\rho$} \|^2  \nonumber \\
&=&\sum_{i=1}^t  \left| r_{ii}\hat{a}_i-\left(\rho_i - \sum_{j=i+1}^t r_{ij}\hat{a}_j \right)\right|  ^2 + \nonumber \\
& & \sum_{i=t+1}^r \rho_i^2< C
\label{SD1}
\end{eqnarray}
are in the sphere. Satisfying Eq.\ (\ref{SD1}) is equivalent to
satisfying the following inequalities for all $k=1,\cdots,t$ :
\begin{eqnarray}
\left| r_{kk}\hat{a}_k-\left(\rho_i - \sum_{j=k+1}^t r_{kj}\hat{a}_j \right)\right|^2 < C' - \nonumber \\
  \sum_{i=k+1}^t  \left| r_{ii}\hat{a}_i-\left(\rho_i - \sum_{j=i+1}^t r_{ij}\hat{a}_j \right)\right|^2  \label{SD11}
\end{eqnarray}
where $C'= C-\sum_{i=t+1}^r \rho_i^2$.
SD computes $\mathbf{\hat{a}}$ satisfying Eq.\ (\ref{SD1}) by deciding
$\hat{a}_i$ in order of $i=t,\cdots,1$ from Eq.\ (\ref{SD11}).

To simplify Eq.\ (\ref{SD11}), we define $S_k$, and $D_k$ as 
\begin{equation}
S_k = \left(
\rho_i - \sum_{j=k+1}^t r_{kj}\hat{a}_j \right)/r_{kk}
\end{equation}
\begin{equation}
D_k =
\sum_{i=k+1}^t  \left| r_{ii}\hat{a}_i-\left(\rho_i-\sum_{j=i+1}^t
r_{ij}\hat{a}_j \right)\right|^2 .  
\end{equation}
Then Eq.\ (\ref{SD11})
is written by $S_k$ and $D_k$ as
\begin{equation}
 \left| \hat{a}_i - S_i \right| ^2<(C'-D_i)/|r_{ii}|^2  \label{compl}.
\end{equation}

{
The candidates of $\hat{a}_i$ satisfying Eq.\ (\ref{compl}) are in the
circle with the center $S_i$ and the radius $\sqrt{(C'-D_i)/|r_{ii}|^2}$ on the complex plane. If there is no candidate of $\hat{a}_i$, SD goes back to decision on $\hat{a}_{i+1}$.
If there are some candidates, we must choose one of them. To reduce the
complexity of searching part, a method starting with  $\hat{a}_i$
nearest to $S_i$ among the all candidates is proposed in
\cite{Tornade}. When  SD only treats real numbers, it is clear which  $\hat{a}_i$ is nearest to $S_i$. But
when SD treats complex numbers, finding $\hat{a}_i$ nearest to $S_i$
needs to compute $\left| \hat{a}_i - S_i \right| ^2$ for all $\hat{a}_i
\in \mathbf{S}$ and not necessarily reduces the complexity.
}So we employ another method. In Section 5, SD chooses $\hat{a}_i$ in the increasing order of $|\Re({\hat{a}_i - S_i})|$ and, if there are two or more candidates of $\hat{a}_i$ with the same value of $|\Re({\hat{a}_i - S_i})|$, SD chooses $\hat{a}_i$ with a smaller $|\Im({\hat{a}_i - S_i})|$, where $\Re(\cdot)$ denotes the real part and $\Im(\cdot)$ denotes the imaginary part. When $\Re({\hat{a}_i- S_i})^2$ is larger than
$(C'-D_i)/|r_{ii}|^2$, SD concludes that there is no $\hat{a}_i$ in the circle any more. 

When $\mathbf{\hat{a}}$ satisfies all inequalities (\ref{compl}), SD concludes that
$M\mathbf{\hat{a}}$ is a lattice point in the sphere. Then the
new radius is set to $\|M \mathbf{\hat{a}} - \mathbf{x} \| $ and SD repeats
the same operations until there is no lattice point in the sphere, and
the last point is the closest point. If there is no lattice
point in the sphere with the radius given first, SD will declare the
erasure of signal or increase the radius.

%%%%%%%%%%%%%%%%%%%%%%%%%%%%%%%%%%%%%%%%%%%%%%%%%%%
\section{QR factorization with sort}
\subsection{Changing the order of decisions on $\hat{a}_i$}
In the previous section, we obtained the inequality with each
signal component  
$\hat{a}_i$. In searching part,  the
computational complexity largely depends on the  order of
decisions on $\hat{a}_i$.
In this section, we consider an efficient order of decisions on $\hat{a}_i$.

For a permutation $\sigma$ and $M=(\mathbf{v}_1,\cdots,\mathbf{v}_t)$,
Eq.\ (\ref{multi}) is equivalently described by $P=(\mathbf{v}_{\sigma
(1)},\cdots,\mathbf{v}_{\sigma (t)})$ and $\mathbf{p} = (a_{\sigma
(1)},\cdots, a_{\sigma (t)})^T$ as
\begin{equation}
\mathbf{x}=P \mathbf{p} + \mbox{\boldmath{$\nu$}}.  \label{kousi}
\end{equation}
When SD processes the channel described as Eq.\ (\ref{kousi}), the order of 
decisions on $\hat{a}_i$ follows the order of components of 
$\mathbf{p}$. So we can change the order of decisions on $\hat{a}_i$ arbitrarily by $\sigma$. SD can obtain $\hat{\mathbf{p}}$
that is the ML estimate of $\mathbf{p}$, and one
can get the  ML estimate of the original channel (\ref{multi}) from
$\hat{\mathbf{p}}$ by the inverse permutation $\sigma^{-1}$.

Next we consider the efficient order of decisions on $\hat{a}_i$. The
number of candidates of $\hat{a}_i$ satisfying Eq.\ (\ref{compl}) is proportional to
\begin{equation}
(C'-D_i)/|r_{ii}|^2. \label{size}
\end{equation}
Intuitively we can reduce the complexity of searching part by changing
the order of decisions on $\hat{a}_i$
so that the value of Eq.\ (\ref{size}) is small for large $i$, because $\hat{a}_i$
are decided in order of $i=t,\cdots,1$.
Because the value of Eq.\ (\ref{size})  is inversely proportional to 
$|r_{ii}|^2$, we can
reduce the complexity by constructing the matrix
$R$ so that $|r_{ii}|$ takes the large value for large $i$. 

{
Now we propose QR factorization with sort to compute the efficient order
of decisions on $\hat{a}_i$. QR factorization computes $r_{ii}$ in
increasing order of $i$. QR factorization with sort permutes columns of
the factorized matrix before each computation of $r_{ii}$ such that $r_{ii}$ is minimized.
QR factorization with sort is used for decreasing the error probability of the nulling and canceling decoder in \cite{LST}. In this paper, we use QR factorization with sort for decreasing the complexity of ML decoder without changing the error probability.
}

{
In \cite{Damen}, it is claimed that the order maximizing $\min_{1\leq i
\leq t} |r_{ii}|$ is optimal for reducing the computational complexity
of searching part. But the computation of  this order requires QR
factorizations $t^2/2$ times. In the mobile environment, the fading matrix $M$
often changes. So the computational complexity
of preprocessing part proposed in \cite{Damen} is not negligible because preprocessing
part is computed whenever fading matrix  $M$ changes.
In \cite{Pohst,Damen}, it is also said that we can reduce the complexity of
SD by reordering decisions on $\hat{a}_i$ according to  the norm of
corresponding basis vectors. In Section \ref{section6}, we compare QR
factorization with sort and other methods by computer simulations.
}
%%%%%%%%%%%%%%%%%%%%%%%%%%%%%%%%%%%%%%%%%%%%%%%%%%%%%
\subsection{Algorithm}
\label{section5}
{
In this subsection, we show how QR factorization with sort works.
QR factorization with sort gives a permutation realizing an
efficient order of decisions on $\hat{a}_i$ and QR factorization for the permuted matrix $P$ in Eq.\ (\ref{kousi}). 
The following algorithm is almost the same as \cite{LST}. The method in \cite{LST} is based on Gram-Schmidt algorithm, and our method is based on  Householder method. It is known that Householder method is numerically more stable than Gram-Schmidt algrithm\cite{NUMERICAL}.
}

{
The ordinary QR factorization of $M$ can be sketched as follows: Compute
a unitary matrix $Q_1$ such that the first column of $Q_1M$ is
$(r_{11},0,\cdots,0)^T$. Let $M_2$ be $((r-1) \times (t-1))$ submatrix
of $Q_1M$ with the first column and the first row of $Q_1M$
removed. Compute a unitary matrix $Q_2$ such that the first column of
$Q_2M_2$ is $(r_{22},0,\cdots,0)^T$. The computation
process is recursively repeated until $i = t$. {See \cite{NUMERICAL} for details. }
}

{
We will describe QR factorization with sort. Observe that in the
ordinary QR factorization $r_{11}$ is equal to the norm of the first
column vector of $M$. In order to minimize $r_{11}$, we replace the
first column 
of $M$ with the column with minimum norm. Let $M'$
be the column replaced version of $M$. Compute a unitary matrix $Q_1'$
such that the first column of $Q_1'H'$ is $(r_{11},0,\cdots,0)^T$. Let
$\tilde{M}_2$ be $((r-1)\times ( t-1))$ submatrix of $Q'_1M'$ with the
first column and the first row of $Q'_1M'$ removed. Replace the first
column of $\tilde{M}_2$ with the column with minimum norm in
$\tilde{M}_2$. Let $M'_2$ be the column replaced version of
$\tilde{M}_2$. Compute a unitary matrix $Q'_2$ such that the first
column of $Q'_2H'_2$ is $(r_{2 2},0,\cdots,0)^T$. The computation
process is recursively repeated until $i = t$.
}

{
With this process we get a QR factorization $\hat{Q}\hat{R}$ of the
column permuted matrix $P$ of $M$. If we apply searching part in
Section 2 to $\hat{Q}\hat{R}$, then we get more efficiently the ML
estimate $\mathbf{p}$. The ML estimate of $\mathbf{\hat{a}}$ can be
obtained by the inverse permutation.
}
%%%%%%%%%%%%%%%%%%%%%%%%%%%%%%%%%%%%%%%%%%\%%%%%%%%%
\section{Dijkstra's algorithm}
In this section we apply Dijkstra's algorithm to searching part to reduce the complexity of searching part  with increase in the storage complexity. 
Dijkstra's algorithm is an efficient algorithm to find the shortest path
from a point to a destination in a weighted directed graph \cite{Dijk}. In this algorithm,
the vertices on the graph are searched for in order of their distance from
the departure. 

The decisions on $\hat{a}_i$ essentially constructs a tree where  
nodes at $k$-th level are correspond to the  candidates of
$\hat{a}_{t-k+1}$ \cite{radius}, and the root is placed at the $0$-th level. Set the
weight of the branch from the node $\hat{a}_i$ to its parent to    
\begin{equation}
\left| r_{ii}\hat{a}_i-\left(\rho_i-\sum_{j=i+1}^t
r_{i,j}\hat{a}_j \right)\right|^2 = r_{ii}^2| \hat{a}_i - S_i |^2.
\end{equation}
Then the distance of node  $\hat{a}_i$ from the root is equal to $D_{i-1}$.    
The nodes having the same parent are arranged in the increasing order of
the distance from left to right. 

If we use Dijkstra's algorithm to find the shortest path from the root
to one of nodes at the bottom level, we can get the node with the minimum $D_0 = \|
\mathbf{x} - M \mathbf{\hat{a}} \|^2$ among all nodes at the bottom level
and it corresponds to the ML estimate.

We show Dijkstra's algorithm.
\begin{enumerate}
\item Create an empty priority queue for nodes.
      The priority is the distance from the root.
\item Insert the leftmost node at the  first level into the priority queue.
\item Select the node $A$ having smallest distance in the priority queue
      and remove it from the  priority queue. If the level of $A$ is $t$, finish this algorithm.
\item Insert the leftmost $A$'s child node into the priority queue.
\item Insert the right neighboring node of $A$ into the priority queue.
\item Go back to Step 3
\end{enumerate}
Because the
node selected in Step 3 has the smaller distance than the nodes selected later, the node at the bottom level selected first
has the minimum value of $D_0$ among all nodes at the bottom level.

In the sequel, we refer to SD not using  Dijkstra's
algorithm as SD, and SD using Dijkstra's
algorithm as Dijkstra's algorithm.

Figure 1 shows an example of the order of search by  Dijkstra's
algorithm and SD. The values in circles show the distance from the root. The numbers in upper rectangles show the order by Dijkstra's algorithm
and the numbers in lower rectangles show the order by 
SD. SD is the depth first search
algorithm for a tree. In this case, the
number of searched nodes  is 5 by Dijkstra's algorithm and is 8
by SD.

 Dijkstra's algorithm
searches for only the nodes whose distance is smaller than the minimum distance of nodes at the bottom level,
but SD searches for the node whose  distance is smaller than $C$ and $C$ must be
greater than the minimum distance of nodes at the bottom level in
order for ML detection succeed. So
the number of searched nodes of Dijkstra's algorithm is smaller than that of SD.
However because we use the priority queue in Dijkstra's algorithm, the storage
complexity increases.
In addition, Dikstra's algorithm does not require the radius of the
sphere to be initially set, and it always finds out ML estimate without
retrying  to search for a lattice point with increased radius.

\begin{figure}[h]
\begin{center}
\label{tree}
\includegraphics[width=0.98\linewidth]{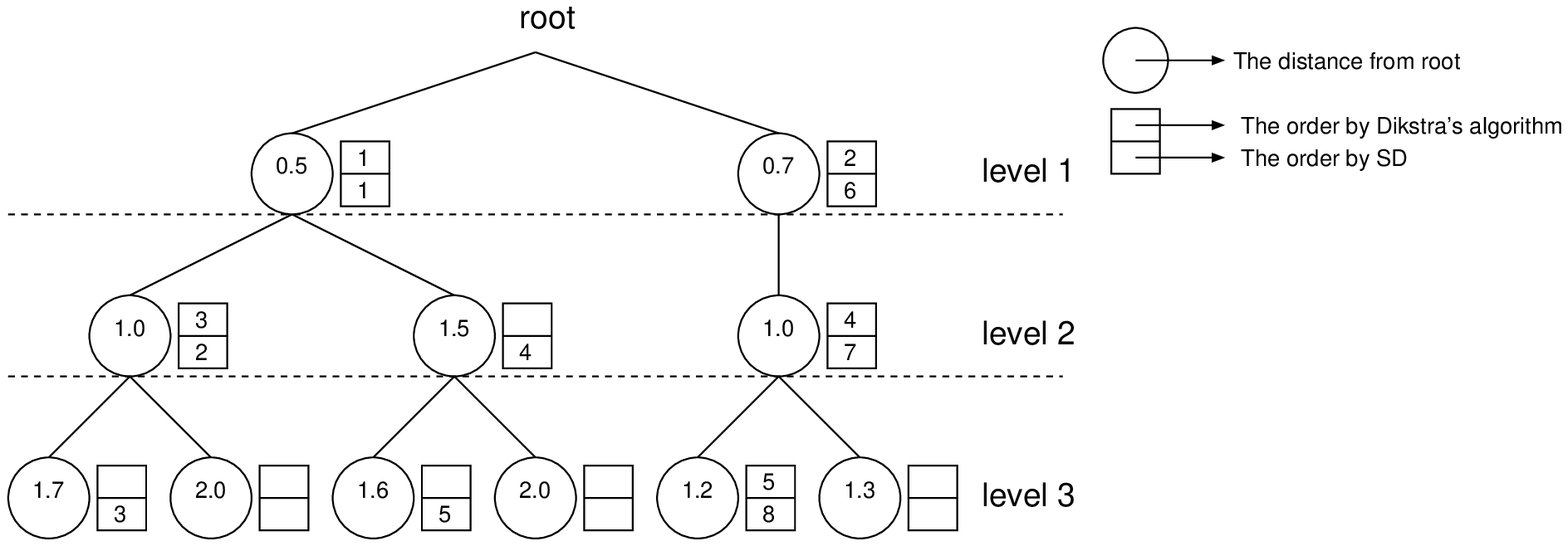}
\caption{The order of search by Dijkstra's algorithm and the original SD}
\end{center}
\end{figure}

Arranging the nodes having the same parent according to the distance
needs to compute $| \hat{a}_i - S_i |^2$ of nodes. Instead of doing
this,  our algorithm considers the candidates of
$\Re(\hat{a}_{t-k+1})$ and the candidates of $\Im(\hat{a}_{t-k+1})$
separately in Section 5. Then the level of tree is equal to $2t$
excluding the root, and
arranging the nodes having the same parent only needs to compute $|
\Re(\hat{a}_i) - \Re(S_i) |$ and $| \Im(\hat{a}_i) - \Im(S_i) |$

\section{Computer simulation} 
\label{section6}
In this section, we show how much the complexity of searching part is
reduced by QR factorization with sort  and Dijkstra's algorithm,
the complexity of preprocessing part is increased by QR factorization
with sort,  and 
the storage complexity is increased by Dijkstra's algorithm over an  uncoded multi-antenna fading channel. 

The radius of sphere used by SD is defined so that 
\begin{eqnarray}
\Pr \{\mbox{transmit point is in sphere}\}&=& \Pr\{C> \left| \mbox{\boldmath{$\nu$}} \right| ^2 \}\nonumber \\ 
&\approx &0.99
\end{eqnarray}
where $C$ is the square of radius and {\boldmath{$\nu$}} is a
vector whose element is noise at each receive antenna
 \cite{radius}.
When there is no lattice point in sphere, we increase the radius to $C+1$,
and continue until a lattice point is found.  
\subsection{The system model}
We consider the following system model.
\begin{itemize}
\item The number of transmit antennas is equal to the number of receive antennas.
\item The fading coefficients obey the  ${\cal CN}(0,1)$ distribution.
\item The signal constellation for each transmit antenna is $64$-QAM
       and all signals are drawn according to the uniform i.i.d. distribution.
%\item The value of SNR at each receive antenna is set to 28dB. 
\end{itemize}

\subsection{The computer simulations}
{
In this subsection we show comparisons of complexities of the proposed
methods and other variants of SD. We remark that all methods in this
subsection are ML decoding and hence the error rates of these ML decoding methods are the same. 
}
First we show the comparison of the complexities of SD not reordering decisions on $\hat{a}_i$
 (SD),  SD reordering decisions on $\hat{a}_i$ according to norms
of basis vectors (Norm-SD) \cite{Pohst,Damen}, SD reordering decisions
 on $\hat{a}_i$ so that $\min_{1 \leq i \leq t} |r_{ii}|$ is maximized (Optimal-SD) \cite{Damen} and SD with
the QR factorization with sort {(QR sort-SD)}. The value of SNR is set to 26dB.
In these simulations
we use the average number of real multiplications and divisions for each processing
as the measure of complexity, and in these simulations we use the
complex multiplications that needs three real multiplications and seven real
additions, and the
complex divisions that needs five real multiplications, two real divisions, and nine real additions \cite{KNUTH}. Figure 2 shows the complexity of searching
part and Figure 3
shows the complexity of preprocessing part. 
When the number of {transmit antennas}
 is 8 the complexity of searching part
is reduced about 55 percent from the original SD by QR factorization
 with sort. However Figure 3 shows the complexity of preprocessing part
 increases about 10 percent. Figure 4 shows the total complexity of SD for 10
 transmissions with the same fading matrix. In this case the complexity
 of SD is  reduced about 60 percent from the original SD by QR factorization
 with sort.

\begin{figure}[ht]
\includegraphics[width=\linewidth]{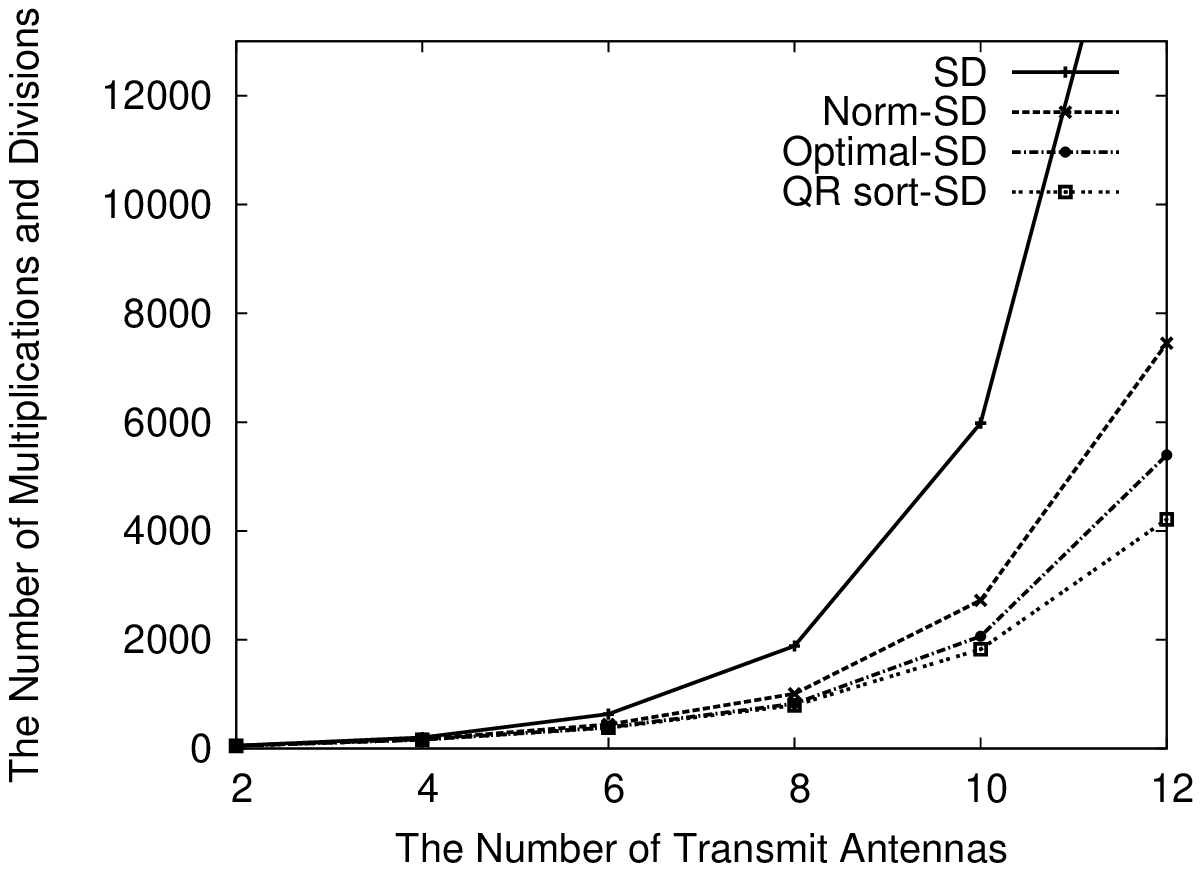}
\caption{The complexity of searching part for each receiving point}
\end{figure}
\begin{figure}[ht]
\includegraphics[width=\linewidth]{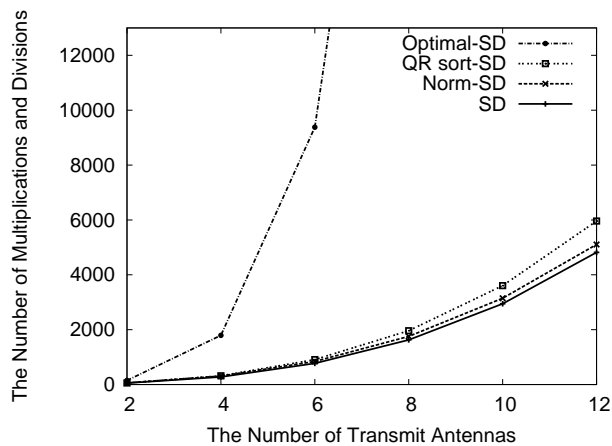}
\caption{The complexity of preprocessing part for each fading matrix}
\end{figure}

\begin{figure}[ht]
\includegraphics[width=\linewidth]{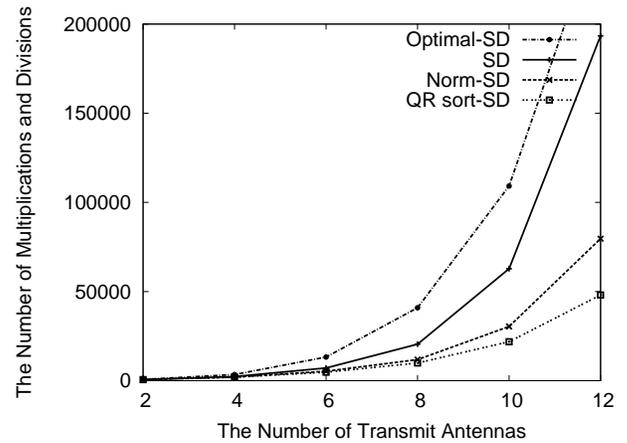}
\caption{The complexity of SD for 10 transmissions  with  the same fading matrix}
\end{figure}

Next we show the comparison of the complexities of  SD (SD), 
Dijkstra's algorithm (Dijkstra), and both of them using QR factorization
with sort {(QR sort-SD, QR sort+Dijkstra)}.

The number of antennas is set to 8.
Figure 5 shows that the complexity of searching part and Figure 6 shows
the cumulative distribution of the size of priority queue with QR
factorization with sort.  When SNR is 26dB, the complexity of searching part
is reduced about 25 percent from the original SD by Dijkstra's
algorithm, and
is reduced about 65 percent from the original SD by combining QR
factorization with sort and Dijkstra's algorithm. Figure 5 also shows
that Dijkstra's algorithm is much faster than SD when SNR is low. 

\begin{figure}[ht]
\includegraphics[width=\linewidth]{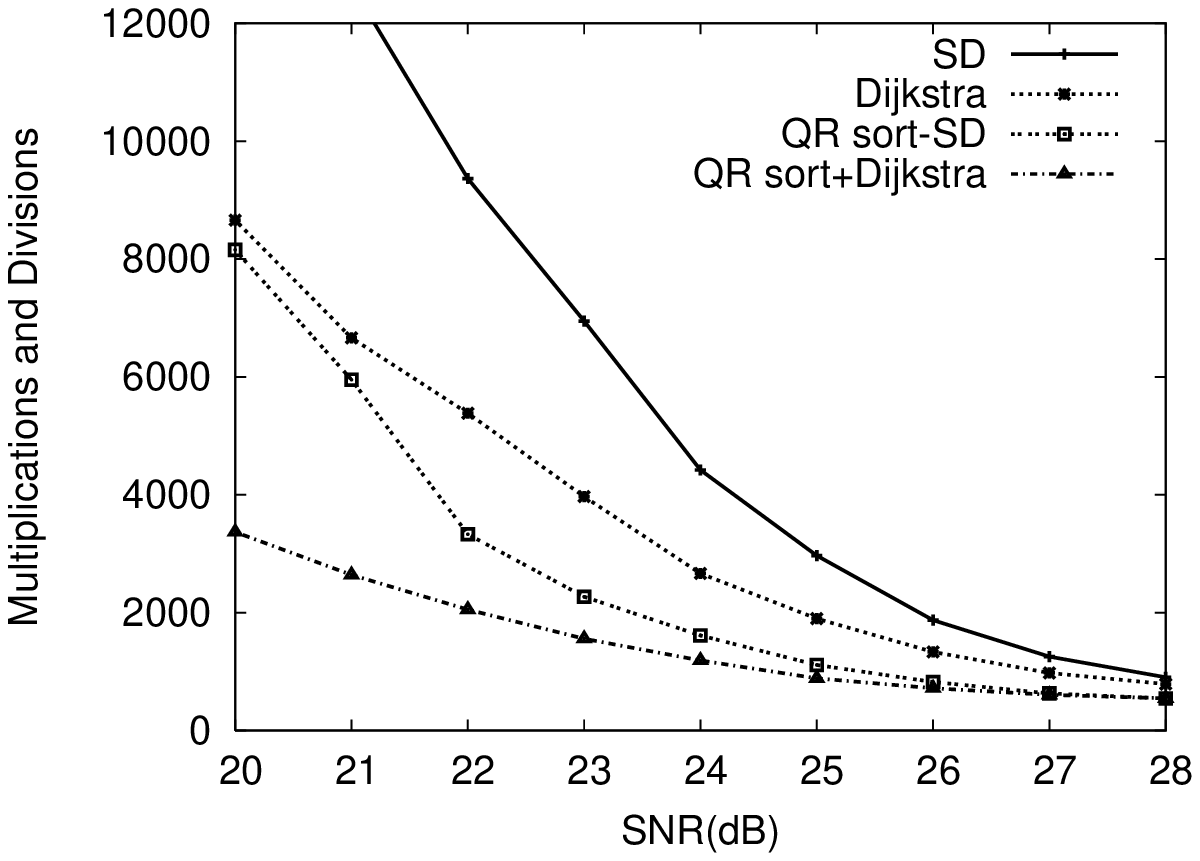}
\caption{The complexity of searching part for each receiving point}
\end{figure}
\begin{figure}[ht]
\includegraphics[width=\linewidth]{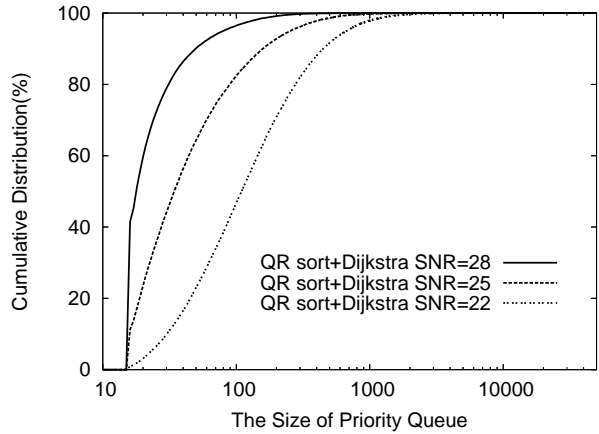}
\caption{The cumulative distribution of the size of priority queue}
\end{figure}

%%%%%%%%%%%%%%%%%%%%%%%%%%%%%%%%%%%%%%%%%
\section{Conclusion}
We proposed the QR factorization with sort and use of Dijkstra's algorithm as
methods for decreasing the computational complexity of the sphere
decoder. QR factorization with sort  reduces the complexity of searching
part of a decoder with little increase in the complexity of preprocessing part of a decoder.
Because the preprocessing part is computed once for each fading
matrix and the increase in the complexity of preprocessing part is little
enough, the total complexity of SD can be reduced. 
Dijkstra's algorithm reduces the complexity of searching part of a decoder with increase in  the storage complexity.
By these reductions of the
complexity, the proposed methods enable us to implement ML decoding for
the multi-antenna system with a lager number of transmit antennas. 
%\bibliographystyle{ieicetr}% bib style
%\bibliography{}% your bib database

\end{document}